\documentclass[11pt]{article}

\usepackage{amssymb,amsmath,amsthm}

\def\R{\mathbb{R}}
\def\eps{\varepsilon}
\def\ra{\rightarrow}
\newtheorem{theorem}{Theorem}[section]
\newtheorem{lemma}{Lemma}[section]

\begin{document}

\title{A Deterministic Polynomial-time Approximation Scheme for Counting Knapsack Solutions}

\author{
Daniel Stefankovic\thanks{
Department of Computer Science, University of Rochester,
Rochester, NY 14627.  Email: stefanko@cs.rochester.edu.
Research supported in part by NSF grant CCF-0910415.
}
\and
Santosh Vempala\thanks{School of Computer Science, Georgia
Institute of Technology, Atlanta GA 30332.
Email: \{vempala,vigoda\}@cc.gatech.edu.
Research supported in part by NSF grant CCF-0830298 and CCF-0910584.}
\and
Eric Vigoda$^\dag$}

\maketitle

\begin{abstract}
Given $n$ elements with nonnegative integer weights $w_1, \ldots,
w_n$ and an integer capacity $C$, we consider the counting version
of the classic knapsack problem: find the number of distinct
subsets whose weights add up to at most the given capacity. We
give a deterministic algorithm that estimates the number of
solutions to within relative error $1\pm\eps$ in time polynomial
in $n$ and $1/\eps$ (fully polynomial approximation scheme). More
precisely, our algorithm takes time $O(n^3\eps^{-1}\log(n/\eps))$.
Our algorithm is based on dynamic programming.  Previously,
randomized polynomial time approximation schemes were known first
by Morris and Sinclair via Markov chain Monte Carlo techniques,
and subsequently by Dyer via dynamic programming and rejection
sampling.
\end{abstract}

\section{Introduction}
Randomized algorithms are usually simpler and faster than their
deterministic counterparts. In spite of this, it is widely
believed that P=BPP (see, e.\,g., \cite{AroraBarak}), i.e., at
least up to polynomial complexity, randomness is not essential.
This conjecture is supported by the fact that there are relatively
few problems for which exact randomized polynomial-time algorithms
exist but deterministic ones are not known. Notable among them is
the problem of testing whether a polynomial is identically zero (a
special case of this, primality testing was open for decades but a
deterministic algorithm is now known,~\cite{AKS}).

However, when one moves to approximation algorithms, there are
many more such examples. The entire field of approximate counting
is based on Markov chain Monte Carlo (MCMC) sampling
\cite{JerrumSinclair:survey}, a technique that is inherently
randomized, and has had remarkable success. The problems of
counting matchings
\cite{JerrumSinclair:matchings,JerrumSinclairVigoda}, colorings
\cite{Jerrum:colorings}, various tilings, partitions and
arrangements \cite{LubyRandallSinclair}, estimating partition
functions \cite{JerrumSinclair:Ising,StefankovicVempalaVigoda}, or
volumes \cite{DyerFriezeKannan,LovaszVempala} are all solved by
first designing a random sampling method and then reducing
counting to repeated sampling.
 In all these cases, when
the input is presented explicitly, it is conceivable that
deterministic polynomial-time algorithms exist.\footnote{Volume
computation has an exponential lower bound for deterministic
algorithms, but that is due to the more general oracle model in
which the input is presented.}

The one notable example of a deterministic approximate counting
algorithm is Weitz's algorithm \cite{Weitz} for counting independent
sets weighted by an activity $\lambda$ for graphs of maximum degree $\Delta$
when $\Delta$ is constant and
$\lambda<\lambda_u(\Delta)$ where $\lambda_u(\Delta)$ is
the uniqueness threshold for the $\Delta$-regular tree.
This was later extended to counting all matchings of bounded degree
graphs \cite{BGKNT}.
An alternative deterministic approach of Bandyopadhyay and
Gamarnik \cite{BG} for colorings and independent sets of bounded degree
graphs only approximates the logarithm of the size of the feasible set.
The results of \cite{Weitz,BGKNT}
are the only two examples of an FPAS (fully polynomial approximation
scheme) for a \#P-complete problem that we are aware of.
One limitation of both of these results is that the running time is quite
large, in particular, the exponent depends on $\ln{\Delta}$.
In contrast, our algorithm has a small polynomial running time.

Here we consider one of the most basic counting problems, namely
approximately counting the number of $0/1$ knapsack solutions.
More precisely, we are given a list of nonnegative integer weights
$w_1, \ldots, w_n$ and an integer capacity $C$,
\footnote{Our results extend to real-valued inputs, but we do not
consider that here to avoid the issue of the model of computation.}
 and wish to count the number of subsets of
the weights that add up to at most $C$. This decision version of
this problem is NP-hard, but has a well-known pseudo-polynomial
algorithm based on dynamic programming. For any $\eps > 0$, we
give a deterministic algorithm that estimates the number of
solutions to within relative error $\eps$ in time polynomial in
$n$ and~$1/\eps$.

Our result follows a line of work in the literature. Dyer et al.
\cite{DFKKPV} gave a randomized subexponential time algorithm for
this problem, based on near-uniform sampling of feasible solutions
by a random walk. Morris and Sinclair \cite{MorrisSinclair}
improved this, showing a rapidly mixing Markov chain, and obtained
an FPRAS (fully polynomial randomized approximation scheme). The
proof of convergence of the Markov chain is based on the technique
of canonical paths and a notion of balanced permutations
introduced in their analysis. In a surprising development, Dyer
\cite{Dyer}, gave a completely different approach, combining
dynamic programming with simple rejection sampling to also obtain
an FPRAS. Although much simpler, randomization still appears to be
essential in his approach---without the sampling part, his
algorithm only gives a factor $n$ approximation.

Our algorithm is also based on dynamic programming, and similar to
Dyer, is inspired by the pseudo-polynomial algorithm for the
decision/optimization version of the knapsack problem. The
complexity of the latter algorithm is $O(nC)$, where $C$ is the
capacity bound. A similar complexity can be achieved for the
counting problem as well using the following recurrence:
\[
S(i,j) = S(i-1,j)+S(i-1,j-w_i)
\]
with appropriate initial conditions. Here $S(i,j)$ is the number
of knapsack solutions that use a subset of the items $\{1,\ldots,
i\}$ and their weights sum to at most $j$.

Roughly speaking, since we are only interested in approximate
counting, Dyer's idea was the following: scale down the capacity
to a polynomial in $n$, scale down the weights by the same factor
and round down the new weights, and then
count the solutions to the new problem efficiently using the
pseudo-polynomial time dynamic programming algorithm. The
new problem could have more solutions (since we rounded down)
but Dyer showed it has at
most a factor of $n$ more for a suitable choice of scaling.
Further, given the exact counting algorithm for the new problem,
one gets an efficient sampler, then uses rejection sampling to
only sample solutions to the original problem. The sampler leads
to a counting algorithm using standard techniques.
Dyer's algorithm has running time $O(n^3 + \eps^{-2}n^2)$ using
the above approach, and $O(n^{2.5}\sqrt{\log(\eps^{-1})} + n^2\eps^{-2})$
using a more
sophisticated approach that also utilizes randomized rounding.

To remove the use of randomness, one might attempt to use a more
coarse-grained dynamic program, namely rather than consider all
integer capacities $1,2, \ldots, C$, what if we only consider
weights that go up in some geometric series? This would allow us
to reduce the table size to $n\log C$ rather than $nC$. The
problem is that varying the capacity even by an exponentially
small factor $(1+n/2^n)$ can change the number of solutions by a
constant factor!

Instead, we index the table by the prefix of items
allowed and {\em the number of solutions}, with the entry in the
table being the minimum capacity that allows these indices to be
feasible. We can now consider approximate numbers of solutions and
obtain a small table.

Our main result is the following:

\begin{theorem}\label{thm:main}
Let $w_1,\dots,w_n$ and $C$ be an instance of a knapsack problem. Let $Z$ be the number
of solutions of the knapsack problem. There is a deterministic algorithm which
for any $\eps\in (0,1)$ outputs $Z'$ such that $(1-\eps) Z\leq Z'\leq Z$.
The algorithm runs in time $O(n^{3}\eps^{-1}\log(n/\eps))$.
\end{theorem}

The running time of our algorithm is competitive with that of
Dyer. One interesting improvement is the dependence on $\epsilon$.
Our algorithm has a linear dependence on $\epsilon^{-1}$ (ignoring
the logarithm term), whereas Monte Carlo approaches, including
Dyer's algorithm \cite{Dyer} and earlier algorithms for this
problem \cite{MorrisSinclair,DFKKPV}, have running time which
depends on $\eps^{-2}$.

\section{Algorithm}
\label{sec:algorithm}

In this section we present our dynamic programming algorithm.
Fix an knapsack instance and fix an ordering on the elements and their weights.

We begin by defining the function
$\tau:\{0,\dots,n\}\times\R_{\geq 0}\ra\R\cup\{\pm\infty\}$
 where $\tau(i,a)$ is the smallest
$C$ such that there exist at least $a$ solutions to the knapsack problem with weights
$w_1,\dots,w_i$ and capacity $C$.
We can not compute the function $\tau$ efficiently since the second argument
ranges over all real numbers.  It will be used in the analysis and it is useful
for motivating the definition of our algorithm.

Note that, by definition, $\tau(i,a)$ is monotone in
$a$, that is,
\begin{equation}\label{monot}
a\leq a' \implies \tau(i,a)\leq\tau(i,a').
\end{equation}
The value of $\tau$ is easy to compute for $i=0$:
\begin{equation}\label{basi}
\tau(0,a)=\Bigg\{\begin{array}{rl}
-\infty & \mbox{if}\ a=0,\\
  0 & \mbox{if}\ 0<a\leq 1,\\
  \infty & \mbox{otherwise}.
\end{array}
\end{equation}
Note that the number of knapsack solutions satisfies:
\begin{equation}\label{number}
Z=\max\{a\, : \, \tau(n,a)\leq C\}.
\end{equation}

We will show that $\tau(i,a)$ satisfies the following recurrence.
\begin{lemma}
\label{lem:recur}
For any $i\in [n]$ and any $a\in\R_{\geq 0}$ we have
\begin{equation}\label{recur}
\tau(i,a)=\min_{\alpha\in [0,1]}
\max\Bigg\{\begin{array}{l}
  \tau\big(i-1,\alpha a),\\
  \tau\big(i-1,(1-\alpha)a\big) + w_i.
\end{array}
\end{equation}
\end{lemma}

We defer the proof of the above lemma to Section \ref{sec:proofs}.

Now we move to an approximation of $\tau$ that we can compute efficiently.
We define a function $T$ which only considers a small set of values $a$
for the second argument in the function $\tau$, these values will form a
geometric progression.

Let
$$
Q:=1+\frac{\eps}{n+1}
$$
and let
$$
s:=\lceil n \log_Q 2\rceil.
$$
The function $T:\{0,\dots,n\}\times\{0,\dots,s\}\ra\R_{\geq 0}\cup\{\infty\}$ is defined
using the recurrence~\eqref{recur} that the function $\tau$ satisfies.
Namely, $T$ is defined by the following recurrence:
$$T[i,j] = \min_{\alpha\in [0,1]} \max\Bigg\{\begin{array}{l}
  T\big[i-1,\lfloor j + \ln_{Q}\alpha\rfloor\big],\\
  T\big[i-1,\lfloor j + \ln_{Q}(1-\alpha)\rfloor\big] + w_i.
\end{array}
$$

More precisely, $T$ is defined by the following algorithm {\bf CountKnapsack}.

\begin{center}
\fbox{\parbox{5in}{
\begin{minipage}{4.5in}
\begin{tt}
{\bf CountKnapsack\\}

Input: Integers $w_1,w_2,\ldots,w_n, C$ and $\eps > 0$.

\begin{enumerate}
\item Set $T[0,0]=0$ and $T[0,j]=\infty$ for $j>0$.
\item Set $Q = (1+\eps/(n+1))$ and $s = \lceil n \log_Q 2\rceil$.
\item For $i=1\rightarrow n$, for $j=0\rightarrow s$, set
\begin{equation}\label{upd:alg}
T[i,j] = \min_{\alpha\in [0,1]} \max\Bigg\{\begin{array}{l}
  T\big[i-1,\lfloor j + \ln_{Q}\alpha\rfloor\big],\\
  T\big[i-1,\lfloor j + \ln_{Q}(1-\alpha)\rfloor\big] + w_i,
\end{array}
\end{equation}
where, by convention, $T[i-1,k] = 0$ for $k<0$.
\item
Let
$$ j' := \max\{j\, : \,T[n,j]\leq C\}.
$$
\item Output $Z' := Q^{j'+1}$.
\end{enumerate}
\end{tt}
\end{minipage}
}}
\end{center}

The minimum in the recurrence \eqref{upd:alg}, although formally over the entire interval $[0,1]$, only needs to be evaluated at the discrete subset where the second argument goes to the next integer.  Hence we will be able to compute $T$ efficiently.

The key fact is that $T$ approximates $\tau$ in the following sense.

\begin{lemma}
\label{lem:approx}
Let $i\geq 1$. Assume that for all $j\in\{0,\dots,s\}$ we have that $T[i-1,j]$ satisfy~\eqref{appr}.
Then for all $j\in\{0,\dots,s\}$ we have that $T[i,j]$ computed using~\eqref{upd:alg}
satisfies:
\begin{equation}\label{appr}
\tau(i,Q^{j-i})\leq T[i,j]\leq \tau(i,Q^j).
\end{equation}
\end{lemma}

We defer the proof of Lemma \ref{lem:approx} to Section \ref{sec:proofs}.

We can now prove that the output $Z'$ of the
algorithm {\bf CountKnapsack} is a
$(1\pm\eps)$ multiplicative
approximation of $Z$.

Note that $Z'$ is never an underestimate of $Z$, since,
\[
C < T[n,j'+1]\leq\tau(n,Q^{j'+1}),
\]
that is, there are at most $Q^{j'+1}$ solutions. We also have
\[
\tau(n,Q^{j'-n})\leq T[n,j']\leq C,
\] that is, there are at least $Q^{j'-n}$ solutions.
Hence
$$
\frac{Z'}{Z} \leq \frac{Q^{j'+1}}{Q^{j'-n}} = Q^{n+1}\leq {\mathrm e}^{\eps}.
$$

This proves that the output $Z'$ of the algorithm {\bf CountKnapsack}
satisfies the conclusion of Theorem \ref{thm:main}.
It remains to show that the algorithm can be modified to achieve the
claimed running time.

\subsection{Running Time}

As noted earlier, the minimum in the
recurrence \eqref{upd:alg} only needs to be evaluated at the discrete subset $S$
where the second argument goes to the next integer.
For $j\in \{0,1,\dots,s\}$,
the set $S$ is $S=S_1\cup S_2$ where:
\[
S_1 = \{Q^{-j},\dots,Q^{0}\} \mbox{ and }
S_2 = \{1-Q^0,\dots,1-Q^{-j}\}.
\]
Thus, $T[i,j]$ can be computed in $O(s)$ time.  Since
there are $O(ns)$ entries of the table and $s=O(n^2/\eps)$ the
algorithm {\bf CountKnapsack} can be implemented in $O(ns^2) = O(n^5/\eps^2)$ time.

To improve the running time, recall that $\tau(i,a)$ is a non-decreasing function in $a$.
Similarly, it is easy to see by induction that $T[i,j]$ is a non-decreasing
function in $j$.  Hence, in \eqref{upd:alg}, the first argument in the
maximum (namely, $T\big[i-1,\lfloor j + \ln_{Q}\alpha\rfloor\big]$)
is non-decreasing in $\alpha$.
Similarly, the second argument in the maximum is a non-increasing function in $\alpha$.
Hence
the minimum of the maximum of the two arguments occurs either at
the boundary (that is, for $\alpha\in\{0,1\}$) or for $\alpha\in
(0,1)$ where the derivative changes from negative to positive,
that is $\alpha$ such that for $\beta<\alpha$
$$T\big[i-1,\lfloor j + \ln_{Q}\beta\rfloor\big] < T\big[i-1,\lfloor j + \ln_{Q}(1-\beta)\rfloor\big] + w_i,$$
and for $\beta>\alpha$
$$T\big[i-1,\lfloor j + \ln_{Q}\beta\rfloor\big]\geq T\big[i-1,\lfloor j + \ln_{Q}(1-\beta)\rfloor\big] + w_i.$$
Therefore, if we had the set $S$ in sorted order, we can find the
$\alpha$ that achieves the minimum in \eqref{upd:alg} using binary
search in $O(\log{s})$ time.  We do not have $S$ in sorted order,
but we do have $S_1$ and $S_2$ in sorted order. We can instead do
binary search over $S_1$ to find the $\alpha\in S_1$ that achieves
the minimum over that set, and then over $S_2$, and finally
compare the two values. Therefore, step 3 of the algorithm {\bf
CountKnapsack} to compute $T[i,j]$ can be implemented in
$O(\log{s})$ time, and the entire algorithm then takes
$O(n^3\eps^{-1}\log(n/\eps))$ time. This completes the proof of
the running time claimed in Theorem \ref{thm:main}.

\section{Proofs of Lemmas}
\label{sec:proofs}

Here we present the proofs of the earlier lemmas.

We begin with the proof of Lemma \ref{lem:recur} which presents the
recurrence for the function $\tau(i,a)$.

\begin{proof}[Proof of Lemma \ref{lem:recur}]
Fix any $\alpha\in [0,1]$. Let $B=\max\{\tau\big(i-1,\alpha
a),\tau\big(i-1,(1-\alpha)a\big) + w_i\}$. There exist at least
$\alpha a$ solutions with weights $w_1,\dots,w_{i-1}$ and capacity
$B\geq \tau\big(i-1,\alpha a)$. There exist at least $(1-\alpha)a$
solutions with weights $w_1,\dots,w_{i-1}$ and capacity $B-w_i\geq
\tau\big(i-1,(1-\alpha) a)$. Hence there exist at least $a$
solutions with weights $w_1,\dots,w_i$ and capacity $B$ and thus
$\tau(i,a)\leq B$. To see that we did not double count, note
that the first type of solutions (of which
there are at least $\alpha a$) has $x_i=0$ and the second type of
solutions (of which there are at least $(1-\alpha)a$) has $x_i=1$.

We established
\begin{equation}\label{elo}
\tau(i,a)\leq \min_{\alpha\in [0,1]}
\max\Bigg\{\begin{array}{l}
  \tau\big(i-1,\alpha a),\\
  \tau\big(i-1,(1-\alpha)a\big) + w_i.
\end{array}
\end{equation}

Consider the solution of the knapsack problem with weights $w_1,\dots,w_i$ and capacity $C=\tau(i,a)$
that has at least $a$ solutions. Let $\beta$ be the fraction of the solutions that do
not include item $i$. Then $\tau(i-1,\beta a)\leq C$, $\tau(i,(1-\beta)a)\leq C-w_i$, and
hence
$$
\max\{\tau(i-1,\beta a),\tau(i,(1-\beta)a)+w_i\}\leq C =
\tau(i,a).
$$
We established
\begin{equation}\label{eup}
\tau(i,a)\geq \min_{\alpha\in [0,1]}
\max\Bigg\{\begin{array}{l}
  \tau\big(i-1,\alpha a),\\
  \tau\big(i-1,(1-\alpha)a\big) + w_i.
\end{array}
\end{equation}
Equations~\eqref{elo} and~\eqref{eup} yield~\eqref{recur}.
\end{proof}

We now prove Lemma \ref{lem:approx} that the function $T$ approximates $\tau$.
\begin{proof}[Proof of Lemma \ref{lem:approx}]
By the assumption of the lemma and~\eqref{monot} we have
\begin{equation}\label{h1}
T\big[i-1,\lfloor j + \ln_{Q}\alpha\rfloor\big]\geq\tau(i-1,Q^{\lfloor j + \ln_{Q}\alpha\rfloor-(i-1)})
\geq\tau(i-1,\alpha Q^{j-i}).
\end{equation}
and
\begin{multline}\label{h2}
T\big[i-1,\lfloor j + \ln_{Q} (1-\alpha)\rfloor\big]\geq\tau\left(i-1,Q^{\lfloor j + \ln_{Q}(1-\alpha)\rfloor-(i-1)}\right)
\\
\geq\tau(i-1,(1-\alpha) Q^{j-i}).
\end{multline}
Combining~\eqref{h1} and~\eqref{h2} with min and max operators we obtain
\begin{equation*}
\begin{split}
\Bigg(\min_{\alpha\in [0,1]} \max\Bigg\{\begin{array}{l}
  T\big[i-1,\lfloor j + \ln_{Q}\alpha\rfloor\big],\\
  T\big[i-1,\lfloor j + \ln_{Q}(1-\alpha)\rfloor\big] + w_i
\end{array}\Bigg) \geq \\
\Bigg(\min_{\alpha\in [0,1]} \max\Bigg\{\begin{array}{l}
  \tau(i-1,\alpha Q^{j-i}),\\
  \tau(i-1,(1-\alpha) Q^{j-i}) + w_i
\end{array}\Bigg) = \tau(i,Q^{j-i}),
\end{split}
\end{equation*}
establishing that $T[i,j]$ computed using~\eqref{upd:alg} satisfy the lower bound in~\eqref{appr}.

By the assumption of the lemma and~\eqref{monot} we have
\begin{equation}\label{h3}
T\big[i-1,\lfloor j + \ln_{Q}\alpha\rfloor\big]\leq\tau(i-1,Q^{\lfloor j + \ln_{Q}\alpha\rfloor})
\leq\tau(i-1,\alpha Q^{j}).
\end{equation}
and
\begin{equation}\label{h4}
T\big[i-1,\lfloor j + \ln_{Q} (1-\alpha)\rfloor\big]\leq\tau(i-1,Q^{\lfloor j + \ln_{Q}(1-\alpha)\rfloor})
\leq\tau(i-1,(1-\alpha) Q^{j}).
\end{equation}
Combining~\eqref{h3} and~\eqref{h4} with min and max operators we obtain
\begin{equation*}
\begin{split}
\Bigg(\min_{\alpha\in [0,1]} \max\Bigg\{\begin{array}{l}
  T\big[i-1,\lfloor j + \ln_{Q}\alpha\rfloor\big],\\
  T\big[i-1,\lfloor j + \ln_{Q}(1-\alpha)\rfloor\big] + w_i
\end{array}\Bigg) \leq \\
\Bigg(\min_{\alpha\in [0,1]} \max\Bigg\{\begin{array}{l}
  \tau(i-1,\alpha Q^{j}),\\
  \tau(i-1,(1-\alpha) Q^{j}) + w_i
\end{array}\Bigg) = \tau(i,Q^{j}),
\end{split}
\end{equation*}
establishing that $T[i,j]$ computed using~\eqref{upd:alg} satisfy the upper bound in~\eqref{appr}.
\end{proof}


%

%

\bibliographystyle{abbrv}
\bibliography{knapsack}

\end{document}